\documentclass[12pt]{article}
\usepackage[hmargin=2.5cm,vmargin=3cm]{geometry}
 \usepackage{graphicx}
 \usepackage{amsmath}
 \usepackage{amssymb}
  \usepackage{enumerate}

\begin{document} \title{Path of a tunneling particle} \author {Charis Anastopoulos\footnote{anastop@physics.upatras.gr} and   Ntina Savvidou\footnote{ksavvidou@upatras.gr}\\
 {\small Department of Physics, University of Patras, 26500 Greece} }

\maketitle
\begin{abstract}
Attempts to find a quantum-to-classical correspondence in a classically forbidden region leads to non-physical paths, involving, for example,  complex time or spatial coordinates. Here, we identify genuine quasi-classical paths for tunneling in terms of probabilistic correlations in sequential time-of-arrival measurements. In particular, we
 construct  the {\em post-selected} probability density $P_{p.s.}(x, \tau)$ for a particle to be found at time $\tau$ in position $x$ inside the forbidden region, provided that it later crossed the barrier. The classical paths follow from the maximization of  the probability density with respect to $\tau$. For almost monochromatic initial states, the paths correspond to the maxima of the modulus square of the wave-function $|\psi(x,\tau)|^2$ {\em with respect to $\tau$} and for constant $x$ inside the barrier region. The derived paths are expressed in terms of classical equations, but they have no analogues in classical mechanics. Finally, we evaluate the paths explicitly for the case of a square potential barrier.
\end{abstract}

\section{Introduction}
Quantum tunneling highlights a key difference between classical and quantum physics. While a classical system may be restricted in a region of the classical state space, the corresponding quantum system explores the totality of the state space, traversing even the classically forbidden regions.The  quantum-to-classical correspondence in tunneling systems is restricted to the classically allowed regions. Looking for such a correspondence in the forbidden regions leads to `paths' that involve either complex time or complex spatial coordinate, i.e.,  no physical paths.

In this article, we identify actual classical paths that provide an approximate quasi-classical description of tunneling. We achieve this by constructing the {\em post-selected} probability density $P_{p.s.}(x, \tau)$ for a particle to be found at time $\tau$ in position $x$, provided that it was detected on the other side of the barrier at some later time.  The probability density is constructed using the Quantum Temporal Probabilities (QTP) description of tunneling \cite{AnSav13}.
We identify tunneling paths  from the maxima of the probability density $P_{p.s.}(x,\tau)$.

The simplest   correspondence between classical paths and quantum evolution follows from
the Wentzel-Kramers-Brillouin (WKB) method \cite{Merz}:  the classical Hamilton-Jacobi equation emerges from an approximate solution to  Schr\"odinger's equation. However, this  correspondence   fails at the classically forbidden region. In Feynman's path integral reformulation of quantum mechanics, tunneling is  described in terms of {\em Euclidean instantons} \cite{instantons}, i.e., `paths' defined with respect to {\em imaginary} time. Alternatively, one can express tunneling in terms of complex-time paths \cite{AoHa}, or in terms of complex-valued solutions to the classical equations of motion \cite{Bender, Turok, Dunne}. However, none of the paths above is meaningful as a path in the phase space of the classical system or is compatible with a spacetime description.

One consequence of the lack of classical-quantum correspondence in tunneling  is an ambiguity in the definition of tunneling time, i.e.,  of the time that it takes a quantum particle   to tunnel through the classically forbidden region.
The search for an answer to this question originates from the early days of quantum mechanics \cite{Con, McC} and it has led     to several different candidates for the tunneling time, rather than to a single expression derived unambiguously from first principles---for reviews, see Ref. \cite{rev1, Raz}.

In this paper, we derive the paths associated to tunneling, by following a procedure  for deriving classical equations for quantum systems that has been outlined by Gell-Mann and Hartle \cite{GeHa}. Classical equations are defined in terms of {\em probabilistic  correlations} between measurement records  obtained at different moments of time. Such equations are classical in the sense that they are deterministic, for example,  differential or integro-differential equations. However, they may not be classical  in the sense of corresponding to classical physics. For this reason, they are   denoted as {\em quasi-classical}. It turns out that the quasi-classical paths for tunneling systems that we derive in this paper have no  analogue   in classical mechanics.

 We find the quasi-classical equations for  tunneling particles by constructing the post-selected   probability density $P_{p.s.}(x, \tau)$ that was referred to earlier, and maximizing it with respect to $\tau$. It is important to note that the position $x$ in $P_{p.s.}(x, \tau)$  is a {\em fixed parameter} designating the location of the detector in the classically forbidden region, and the time $\tau$ is a random variable. Thus,  $P_{p.s.}(x, \tau)$ is maximized with respect to $\tau$, while keeping  $x$ fixed. The  more familiar case of maximizing probabilities with respect to $x$ for fixed $\tau$ leads to values of $x$ outside the classically forbidden region, and thus, cannot lead to effective equations for tunneling.

 Thus, the key point in our approach is that the time $\tau$ is  treated as a random variable because the  post-selected   probability density $P_{p.s.}(x, \tau)$ is defined in terms of {\em time-of-arrival measurements}, and not the usual von Neumann measurements. To this end, we employ  the QTP description of tunneling systems \cite{AnSav13}, generalized for  sequential measurements \cite{AnSav16}. The probability density $P_{p.s.}(x, \tau)$ is  a linear functional of a four-point function of the associated quantum fields.

The QTP method provides a general procedure for constructing probabilities associated to temporal observables, i.e., time variables whose value can be determined in specific experiments. It was first developed in order to address the time-of-arrival problem \cite{AnSav12, AnSav06}, and it has been applies to the temporal description of tunneling \cite{AnSav13, AnSav08}, non exponential decays \cite{An08} and to relativistic quantum measurements  \cite{AnSav12, AnSav11, AnSav15}.
The key idea is to distinguish between the roles of time as a parameter to Schr\"odinger’s equation and as a label of the causal ordering of events \cite{Sav99, Sav10}. This important distinction leads to the definition of quantum temporal observables. In particular, we identify the time of a detection event as a coarse-grained quasi-classical variable  \cite{GeHa}  associated with macroscopic records of observation. The time variables   correspond to macroscopic observable magnitudes, such as the coincidence of a detector `click’ with the reading of a clock external to the system.

The quasi-classical equations for tunneling  turn out to be particularly  simple for  initial states that are well localized in momentum. They are obtained from the maximization of the modulus square $|\psi(\tau, x)|^2$ of the wave function {\em with respect to $\tau$} and for fixed $x$ inside the classically forbidden region. Maximization leads to a functional relation for the detection time $\tau$ as a function of the position $x$: $\tau = \tau(x)$. If the function $\tau(x)$ is  invertible, it defines a path $x(\tau)$ in the classical forbidden region. This is   the case for tunneling in the square potential barrier that is analytically tractable.  Interestingly, the WKB approximation turns out to be insufficient for the evaluation of the  quasi-classical equations derived here.

\medskip

The structure of this paper is the following. In Sec. 2, we summarize the QTP description  tunneling. In Sec. 3, we derive the post-selected   probability density $P_{p.s.}(x, \tau)$. In Sec. 4, we derive the quasi-classical equations   and specialize to the case of the square potential barrier where the solutions to the equations are exact. In Sec. 5, we summarize and discuss our results. In the Appendix, we elaborate on the  probabilities associated to sequential time-of-arrival measurements and we discuss the WKB approximation.

\section{Background}
\subsection{QTP time-of-arrival probabilities}
In this section, we present  the main results of the QTP method for tunneling Ref. \cite{AnSav13}, also setting up the notation to be used in later sections.

Let ${\cal H}$ be the Hilbert space associated to a particle in one dimension, described by a Hamiltonian $\hat{h}$. We denote the generalized eigenstates of the Hamiltonian as    $|a\rangle$ and the eigenfunctions as $f_a(x)$.

The QTP method requires the introduction of quantum fields even for non-relativistic particles, so we extend the description on the Fock space ${\cal F}$ associated to ${\cal H}$.
 For the purposes of this paper, it makes no difference whether the particles are bosons or fermions, so we arbitrarily choose fermions. The annihilation and creation operators on ${\cal F}$,    $\hat{c}_a$ and $\hat{c}_a^{\dagger}$ respectively,  satisfy the canonical anticommutation relations
\begin{eqnarray}
\{\hat{c}_a, \hat{c}_b\} = \{\hat{c}^{\dagger}_a, \hat{c}^{\dagger}_b\} = 0, \hspace{1.3cm} \{\hat{c}_a, \hat{c}^{\dagger}_b\} = \delta_{ab}.
\end{eqnarray}
 The Heisenberg picture field operators are defined as
 \begin{eqnarray}
 \hat{\psi}(x, t) = \sum_a f_a(x) e^{-i\epsilon_at} \hat{c}_a, \hspace{1cm}  \hat{\psi}^{\dagger}(x, t) = \sum_a f^*_a(x) e^{i\epsilon_at} \hat{c}^{\dagger}_a.
 \end{eqnarray}
 We will denote the vacuum state of ${\cal F}$ by $|0\rangle$.

  Given an initial state $|\Psi\rangle \in {\cal F}$,  the probability that a detector located at $x = L$ records a particle within the time interval $[t, t+dt]$ is $P(L,t)dt$, where  $P(L, t)$ is  a linear functional of 2-pt correlation function \cite{AnSav12, AnSav15}
  \begin{eqnarray}
  P(L, t) = C \int ds g(s) \langle \Psi_0|\hat{Y}^{\dagger}(L,t-\frac{s}{2})\hat{Y}(L, t+\frac{s}{2})|\Psi_0\rangle. \label{plt1}
  \end{eqnarray}
  In Eq. (\ref{plt1}), $C$ is a normalization constant; $g(s)$ is an even  function that is peaked around $s = 0$ and depends upon the detailed properties of the detector.

  The composite  operator $\hat{Y}({\pmb x}, t)$ is   a local functional of the quantum fields $\hat{\psi}(x)$ and $\hat{\psi}^{\dagger}(x)$ and originates from the interaction Hamiltonian between the particles and the detector. In what follows,   two forms of $\hat{Y}(x)$ are relevant,
\begin{enumerate}
\item $\hat{Y}(x, t) = \hat{\psi}(x, t)$. It describes  a process in which the particle is absorbed during detection.
    \item $\hat{Y}( x, t) = \hat{\psi}^{\dagger}(x, t)\hat{\psi}(x, t)$. It describes a process in which the particle is scattered during detection.
\end{enumerate}

Choosing the case (i) above, of detection by absorption, and for a single-particle state $\hat{\Psi}_0 = \sum_a \psi_{0a}\hat{c}^{\dagger}_a|0\rangle$, Eq. (\ref{plt1}) becomes

\begin{eqnarray}
P(L, t) = C  \sum_{a,a'} \tilde{g}\left[\frac{1}{2} (\epsilon_a+\epsilon_{a'})\right]   f_a(L) f^*_{a'}(L) e^{- i (\epsilon_a - \epsilon_{a'}) t} \psi_{0a}\psi^*_{0a'}, \label{plt2}
\end{eqnarray}
where $\tilde{g}$ is the Fourier transform of $g(s)$.

\subsection{Time-of-arrival probabilities in a tunneling system}
 Next, we specialize to a set-up relevant to tunneling. We consider  a single-particle Hamiltonian
 \begin{eqnarray}
 \hat{h} = \frac{\hat{p}^2}{2m} + V(\hat{x}),
 \end{eqnarray}
 where $V(x)$ is a potential that vanishes for  $x \notin [ -a, a]$, it is everywhere non-negative, and it is parity-symmetric: $V(x) = V(-x)$. The last two  assumptions are not essential, but they remove technical complications that are peripheral to the main aims of this paper.

The Hamiltonian $\hat{h}$
   has double degeneracy for each positive value of energy $\epsilon$. We  denote the generalized eigenvectors as $|k, \pm\rangle$ where  $k = \sqrt{2m\epsilon} \geq 0$, and we write  $f_{k \pm}(x) = \langle x| k, \pm \rangle$. The eigenfunctions $f_{k+}(x)$ behave as $e^{ikx}$ for $x >a$ and the eigenfunctions $f_{k -}(x)$ as $e^{-ikx}$ for $x <-a$.

 In a parity-symmetric potential, $\langle k, +|k,-\rangle = 0$ and $f_{k-}(x) = f_{k +}(-x)$, thus, we only need to specify the functions $f_{k+}$,
\begin{eqnarray}
f_{k+}(x) =  \left\{ \begin{array}{cc} \frac{1}{\sqrt{2\pi}}( e^{ikx} +R_k e^{-ikx}) & x < - a\\
e^{r_k(x) + i \theta_k(x)} & -a \leq x \leq a \\
 \frac{1}{\sqrt{2\pi}}T_k e^{ikx}& x> a\end{array} \right. \label{fe+}
\end{eqnarray}
  $T_k$ is the transmission amplitude and $R_k$ the reflection amplitude for a right-moving particle.

We consider an initial wave function $\psi_0(x)$ localized   to the left ($x <-a$) of the barrier region and with support only on positive momenta. Then   $\langle \psi_0|k, -\rangle \simeq 0$ and $\langle k, +|\psi_0\rangle \simeq \tilde{\psi}_0(k)$, the Fourier transform of $\psi_0(x)$.   We also assume that the detector is on the other side of the barrier,

\begin{eqnarray}
P(L,t) =  C \int \frac{dk dk'}{2\pi m}T_k T^*_{k'} \tilde{g}[(\epsilon_k+\epsilon_{k'})/2] \tilde{\psi}_0(k) \tilde{\psi}_0^*(k')   e^{i(k-k')L-i(\epsilon_k-\epsilon_{k'})t}, \label{probden}
\end{eqnarray}
where   $\epsilon_k = \frac{k^2}{2m}$.

Eq. (\ref{probden}) is physically meaningful for $t \geq 0$ but is also defined   for $t <0$. With the initial state $\psi_0(x)$ considered here,  the values of $P(L,t)$ for $t<0$ are strongly suppressed. Hence, we can evaluate the total probability of detection $P_{tot}(L) $ by integrating $t$  over the full real axis, to obtain
\begin{eqnarray}
P_{tot} = Cm  \int_0^{\infty} dk |T_k|^2 \frac{\tilde{g}(\epsilon_k)}{k} |\tilde{\psi}_0(k)|^2.
\end{eqnarray}

As can be seen by examining the case of $V(x) = 0$,  the quantity $\tilde{g}(\epsilon_k)/k$ is proportional to the absorption rate $\alpha(k)$ of the detector (number of detected particles modulo number of incoming particles momentum $p$). Assuming an ideal detector, with the same absorption rate for all momenta, $\tilde{g}(\epsilon_k) \sim  k = \sqrt{2m \epsilon_k}$. We  choose the proportionality coefficient so that $P_{tot}(L) = 1$ for $V(x) = 0$ \cite{AnSav13}. Then, we derive the probability density for the time of arrival for tunneling particles
\begin{eqnarray}
P(L,t) =  \int \frac{dk dk'}{2\pi m}  T_k T^*_{k'} \sqrt{\frac{\epsilon_k +\epsilon_{k'}}{m}}\tilde{\psi}_0(k) \tilde{\psi}_0^*(k')  e^{i(k-k')L-i(\epsilon_k-\epsilon_{k'})t}. \label{probden2}
\end{eqnarray}

The total detection probability is then
\begin{eqnarray}
P_{tot}  =   \int_0^{\infty} dk |T_k|^2 |\tilde{\psi}_0(k)|^2, \label{ptot0}
\end{eqnarray}
 and can be expressed as $\langle \psi_0|\hat{\Pi}_+|\psi_0\rangle$, where the positive operator
\begin{eqnarray}
\hat{\Pi}_+ = \int dk |T_k|^2 |k\rangle \langle k|, \label{hatpl}
\end{eqnarray}
 is defined on the subspace of states with positive momentum  and corresponds to  the event of detection.



 \section{ Post-selected detection probability}
In this section, we construct the   probability density associated to a measurement in the forbidden region, post-selected so that only particles that have crossed the barrier are taken into account.  We consider two successive time-of-arrival measurements, one by a detector in the forbidden region and one outside the barrier ($x > a)$.

 We construct the joint probability $P(x, \tau; L,t)$ that  first,  the detector at $x$ will first record a particle at time $\tau$ and then the detector at $L$ will detect the particle at time $t \geq \tau$. It is a linear functional of a four-point function of the associated fields
\begin{eqnarray}
P(x, \tau; L,t) = C \int ds_1 ds_2 g(s_1) g(s_2)  \langle \Psi_0|  \hat{Y}_1^{\dagger}(x ,\tau-\frac{s_1}{2}) \hat{Y}_2^{\dagger}(L, t - \frac{s_2}{2}) \nonumber \\
 \times \hat{Y}_2(L,  t + \frac{s_2}{2}) \hat{Y}_1(x, \tau+\frac{s_1}{2})|\Psi_0\rangle, \label{pltlt2}
\end{eqnarray}
In Eq. (\ref{pltlt2}), $C$ is a normalization constant, the functions $g_{i}(s)$ are  versions of the function $g(s)$ of Eq. (\ref{plt1}) that characterize each detector,     and $\hat{Y}_{i} $ are Heisenberg-picture composite operators that are local functionals of the quantum fields $\hat{\psi}(x, t)$ and $\hat{\psi}^{\dagger}(x, t)$.
For a derivation of Eq. (\ref{pltlt2}), see, Refs. \cite{AnSav15, AnSav16} and for an earlier form, Ref. \cite{AnSav11}. Note that Eq. (\ref{pltlt2}) involves averaging over the temporal and not the spatial coordinates.   Temporal averaging is essential for the definition of probabilities in the QTP method. This is not the case for spatial averaging. It is usually subsumed under the effects of temporal averaging and  can be omitted for simplicity.

A second measurement on a particle is only possible if the first measurement does not annihilate the particle. Hence, the  interaction between the particle and the first detector must describe scattering rather than absorption, $\hat{Y}_1(x)= \hat{\psi}^{\dagger}(x, t)\hat{\psi}(x,t)$.   There is no constraint for the second measurement, so we consider detection by absorption, $\hat{Y}_2(x, t) = \hat{\psi}(x, t)$.

   As in Sec. 2.2,  we consider a single-particle initial state localized to the left of the barrier and with support only on positive momenta. Eq. (\ref{pltlt2}) yields
\begin{eqnarray}
P(x, \tau; L,t) =   \int dk dk' {\cal F}[L, t - \tau; x; \frac{1}{2}(\epsilon_k + \epsilon_{k'})] f_{k+}(x) f_{k'+}(x) e^{-i (\epsilon_k - \epsilon_{k'})\tau} \tilde{\psi}_0(k) \psi_0^*(k'), \label{pdouble}
\end{eqnarray}
where
\begin{eqnarray}
{\cal F}(L, t; x ; E) = C \int dk dk' \tilde{g}_1[\frac{1}{2}(\epsilon_k +\epsilon_k' - 2E)] \tilde{g}_2\left[\frac{1}{2} (\epsilon_k +\epsilon_{k'}  )  \right] e^{-i(\epsilon_k - \epsilon_k')t} \nonumber \\
\times
\left[ f_{k+}(L) f^*_{k+}(x) + f_{k-}(L) f^*_{k-}(x) \right] \left[f^*_{k'+}(L) f_{k+}(x) + f^*_{k'-}(L) f_{k'-}(x) \right], \label{Fker}
\end{eqnarray}
is a kernel containing information about the propagation from the first to the second detector.

  We use Eq. (\ref{fe+}) for   $f_{k\pm}(L)$ in Eq.  (\ref{Fker}). For positive $t$,  integration over $k$ suppresses terms of the form $\exp[\pm i(kL+i \epsilon_k)t]$, and we obtain
  \begin{eqnarray}
  {\cal F}(L, t; x ; E) = C \int \frac{dk dk'}{2\pi} \tilde{g}_1[\frac{1}{2}(\epsilon_k +\epsilon_k' - 2E)] \tilde{g}_2\left[\frac{1}{2} (\epsilon_k +\epsilon_k'  )\right]  e^{i(k-k')L  -i(\epsilon_k - \epsilon_k')t} \nonumber \\
  \times \left[ T_k f^*_{k+}(x) + R_kf^*_{k-}(x)\right] \left[ T^*_{k'} f_{k'+}(x) + R^*_{k'} f_{k'-}(x)\right]. \label{Fker2}
  \end{eqnarray}

We choose functions $\tilde{g}_i$ that correspond to ideal detectors. As in Sec. 2.2, we consider $\tilde{g}_2(\epsilon) \sim \sqrt{\epsilon}$. The function $\tilde{g}_1$ implements energy conservation on the first detector. In the Appendix A, it is shown that
the condition   $\tilde{g}_1 \sim \epsilon_k \delta (\epsilon_k - \epsilon_{k'})$ corresponds to a detector with negligible energy loss due to scattering. We   choose the coefficient $C$, so that
 the total probability $\int dt \int d\tau P(x, \tau; L,t)$ is unity for  free particles \cite{AnSav16}. Then, Eq. (\ref{Fker2}) simplifies
   \begin{eqnarray}
  {\cal F}(L, t; x ; E) =   \int dk dk' \left(\frac{\epsilon_{k}+\epsilon_{k'}}{m}\right)^{3/2} \delta[\frac{1}{2} (\epsilon_k +\epsilon_k') - E]   e^{i(k-k')L -i(\epsilon_k - \epsilon_k')t} \nonumber \\
  \times \left[ T_k f^*_{k+}(x) + R_kf^*_{k-}(x)\right] \left[ T^*_{k'} f_{k'+}(x) + R^*_{k'} f_{k-}(x)\right]. \label{Fker3}
  \end{eqnarray}
The   probability density (\ref{pdouble}) has to be supplemented with probabilities associated to the events that one or the other of the detectors has not recorded particles. This is necessary for the consistent probabilistic description of the system, because the values $t$ and $\tau$ for the time of arrival do not define a complete set of alternatives, unless they are supplemented with the events of no detection. The  probabilities of no detection are described in the Appendix A.

The post-selected probability density $P_{p.s}(x, \tau) := \int_{\tau}^{\infty}  dt  P(x, \tau; L,t) $ is
\begin{eqnarray}
P_{p.s}(x, \tau) =   \int dk dk' {\cal F}^+[x,  \frac{1}{2}(\epsilon_k + \epsilon_{k'})] f_{k+}(x) f_{k'+}(x) e^{-i (\epsilon_k - \epsilon_{k'})t} \tilde{\psi}_0(k) \psi_0^*(k'), \label{pps}
\end{eqnarray}
where
\begin{eqnarray}
{\cal F}^+(x, E) = \int_0^{\infty} dt   {\cal F}(L, t; x ; E). \label{Fkint}
 \end{eqnarray}
 The integral in Eq. (\ref{Fker2}) is strongly suppressed for $t< 0$. Thus, we can extend the range of integration in Eq. (\ref{Fkint}) to $(-\infty, \infty)$, to obtain
 \begin{eqnarray}
 {\cal F}^+(x, E) =  2 \pi \int dk \left(\frac{k}{m}\right)^2\delta(\epsilon_k - E) |T^*_k f_{k+}(x) + R^*_kf_{k-}(x)|^2. \label{Fkint2} \end{eqnarray}

Eqs. (\ref{pps}) and (\ref{Fkint2}) are the main result of this paper. They express the   probability $P_{p.s}(x, \tau)$ that a particle will be found at location $x$ and time $\tau$ in the forbidden region, provided that the particle has been recorded to cross the barrier.

Consider an almost monochromatic initial state $\psi_0(x)$, with mean momentum $k_0$ and momentum spread $\sigma_p << k_0$. In a $\cap$-shaped potential, the transmission and reflection amplitudes do not exhibit periodicity with respect to $k$. There is also no periodicity with respect to $k$ in $f_{k\pm}(x)$ for fixed $x$ in the forbidden region. Thus, for sufficiently small $\sigma_p$, we can approximate
 ${\cal F}^+[x,  \frac{1}{2}(\epsilon_k + \epsilon_{k'})]$ with ${\cal F}^+[x,  \epsilon_{k_0}]$. Hence,  the right-hand of Eq. (\ref{pps}) becomes  ${\cal F}^+[x,  \epsilon_{k_0}] |\psi_{\tau}(x)|^2$, where $\psi_{\tau}(x) = \langle x|e^{-i\hat{h}\tau}|\psi_0\rangle$ is the quantum state at time $\tau$. By Eq. (\ref{Fkint2}),
\begin{eqnarray}
P_{p.s}(x, \tau) \simeq 2 \pi \frac{k_0}{m} |T^*_{k_0} f_{k_0+}(x) + R^*_{k_0}f_{k_0-}(x)|^2 |\psi_{\tau}(x)|^2 \label{pps2}
\end{eqnarray}
The post-selected probability density is maximized when $|\psi_{\tau}(x)|^2$ is maximized for fixed $x$. Hence, the condition
\begin{eqnarray}
\frac{\partial}{\partial \tau} |\psi_{\tau}(x)|^2 = 0 \label{quasicl}
\end{eqnarray}
defines the quasi-classical equations for tunneling.

We note that Eq.  (\ref{quasicl}) only requires  the assumption of an almost monochromatic initial state. The choice of functions $\tilde{g}_i$ in Eq. (\ref{Fker2}) affects the form of the post-selected probability density, but not   the maximum   (\ref{quasicl}).

Solutions of Eq. (\ref{quasicl}) are of the form $\tau(x)$, i.e., the time $\tau$ is determined as a function of $x$. In contrast, a classical path corresponds to a function $x(\tau)$ of time. Thus, the quasiclassical correlations between position and time will correspond to the usual notion of a  path {\em only if the function $\tau(x)$ is bijective}, and hence, can be inverted to define a function $x(\tau)$.

The unusual term $|T^*_k f_{k+}(x) + R^*_kf_{k-}(x)|^2$ in Eq. (\ref{Fkint2}) is a consequence of the choice  $\hat{Y}_1(x)= \hat{\psi}^{\dagger}(x, t)\hat{\psi}(x,t)$ for the composite operator of the first measurement. This operators treats left-moving (+) and right-moving (-) modes symmetrically, so propagation from $x$ to $L$ also involves both left-moving modes that traverse the barrier and right-moving  modes that are reflected upon the barrier. In general, the  composite operator that describes the first measurement should be determined   from a detailed modeling of the measuring apparatus. We believe that the above choice for $\hat{Y}_1$ is the most natural; we find no reason why the interaction with a measurement device in the forbidden region would distinguish between the two types of mode. Of course, other choices for $\hat{Y}_1$ are possible. However, their only effect is to change the form of the $x$-dependent term before $|\psi_{\tau}(x)|^2$ in Eq. (\ref{pps2}).
{\em The quasiclassical equations (\ref{quasicl}) remain the same}.

The maximum (\ref{quasicl}) involves the derivative of $|\psi_{\tau}(x)|^2$ with respect to $\tau$ at constant $x$ and not the derivative with respect to $x$ at constant $\tau$. The reason is that the random variable in Eq. (\ref{pps2}) is $\tau$, while   $x$ is an fixed parameter designating the location of the detector. This feature is essential for the derivation of the quasi-classical equations. To see this, consider an alternative measurement scheme, in which   a von Neumann measurement of position prior to a time-of-arrival measurement after the barrier. In the associated post-selected probability
 $P_{p.s}(x, \tau)$, $\tau$ is a parameter and $x$ is a random variable,
\begin{eqnarray}
 P_{p.s.}(x,\tau) = Tr \left[ \hat{\Pi}^+_L \sqrt{\hat{P}_x}e^{-i\hat{H}\tau} \hat{\rho}_0 e^{i\hat{H}\tau} \sqrt{\hat{P}_x}\right], \label{intprob}
\end{eqnarray}
where $\hat{\rho}_0 = |\psi_0\rangle\langle \psi_0|$, $\hat{\Pi}^+_L$ is given by Eq. (\ref{hatpl}) and $\hat{P}_x$ define a position POVM.

Eq. (\ref{intprob})  has the familiar form of a probability density for a post-selected measurement  \cite{postsel}, and some variations have appeared in the discussion of tunneling \cite{Steinberg, Turok}. However, Eq. (\ref{intprob}) cannot   define quasiclassical equations inside the barrier. The maximization of (\ref{intprob}) with respect to $x$ for constant $\tau$ leads to values of $x$ {\em outside the barrier}. For a localized wave-packet, the maximum of $|\psi_{\tau}(x)|^2$ at fixed $\tau$ evolves approximately according to the classical equations of motion; hence, predicts that the particle is reflected at the classically forbidden region.


In the Appendix A, we   derive the probability density $P_1(x,\tau)$ that the first detector detects a particle at time $\tau$, irrespective of what happens at the second detector,
 \begin{eqnarray}
 P_1(x, \tau) = \int dk dk' \sqrt{\frac{\epsilon_k+\epsilon_{k'}}{m}} f_k(x)f_{k'}(x) e^{-i(\epsilon_k-\epsilon_{k'})\tau} \tilde{\psi}_0(k) \tilde{\psi}^*_0(k') \label{p1xt}
          \end{eqnarray}
For an almost monochromatic initial state $\psi_0(x)$, with mean momentum $k_0$ and momentum spread $\sigma_p << k_0$,
\begin{eqnarray}
 P_1(x, \tau) \simeq \frac{k_0}{m} |\psi(x, \tau)|^2, \label{sdet}
\end{eqnarray}
hence, the corresponding quasiclassical equations are also given by Eq. (\ref{quasicl}). This means that the quasiclassical equations inside the barrier region are largely independent of  the second detection.

 We emphasize that the derivation of the quasi-classical equations from the post-selected probability density $P_{p.s.}(x, \tau)$ is the conceptually correct procedure, because it guarantees that the derived path corresponds to a particle that has actually traversed the barrier.   Eq. (\ref{p1xt}) leads to the same result only because the correlations between the two measurements turns out to be   weak. The weakness of such correlations is a  defining feature of a classically forbidden region; in a classically allowed region, the classical equations of motion lead to strong correlations between the  measurement records.

  Finally, we note that a simple, but non-trivial, consistency check of the formalism can be given by considering the probabilities constructed above at the point $x=a$ where the particle exits the barrier. For the square potential barrier---that is studied in detail in the following section---the point $x = a$ is also the exit of the classically forbidden region. For $x > a$ the particle evolves freely, so we can approximate  $|\psi_{\tau}(x)|^2 \simeq |\psi_{\tau - m(x-a)/k_0}(x)|^2$. Thus, by evaluating $|\psi_{\tau}(a)|^2$ we can obtain the probability density for any position measurement, after the particle has exited the barrier.

 The single-measurement probability distribution (\ref{sdet}) at $x = a$ is
 \begin{eqnarray}
  P_1(a, \tau) \simeq \frac{k_0}{m} |\psi(a, \tau)|^2 =  \frac{k_0}{m}|\psi_{\tau'}(x)|^2,
 \end{eqnarray}
 where $\tau ' = \tau - m(x-a)/k_0$.
Thus, the time-of-arrival probability coincides with the probability density for position $|\psi_{\tau'}(x)|^2$ given by Born's rule, modulo the velocity $k_0/m$ that appears in a change of variables from time to position.

We also evaluate the post-selected probability density (\ref{pps2}) at the exit point $x = a$. By Eq. (\ref{fe+}), $T^*_{k} f_{k+}(a) + R^*_{k}f_{k-}(a) = \frac{1}{\sqrt{2\pi}} (e^{ika} +e^{-ika}R^*_k)$, hence,
\begin{eqnarray}
P_{p.s}(a, \tau) = |1 + e^{-2ik_0a}R_{k_0}|^2\frac{k_0}{m}|\psi_{\tau'}(x)|^2 = |1 + e^{-2ik_0a}R_{k_0}|^2  P_1(a, \tau),
\end{eqnarray}
i.e., it differs from $P_1(a, \tau)$ by a multiplicative factor that is typically of order unity. Thus, quantum post-selection may enhance on diminish the transmission probability through the barrier by a constant factor---the post-selected probability remains proportional to $|\psi_{\tau'}(x)|^2$, as  given by Born's rule.

\section{Quasi-classical equations for tunneling particles}
In this section, we identify the quasi-classical equations of motion for tunneling particles, and we apply the results of Sec. 3 to an exactly solvable model.
\subsection{Maximization of the post-selected probability density}
The quasi-classical equations of motion follow from the maximization condition (\ref{quasicl}). We consider the evolution of an initial state localized at $x_0 < - a$ and  with mean momentum $k_0 > 0$,
\begin{eqnarray}
\psi_0(x) = \phi(x-x_0)  e^{ i k_0(x-x_0)};
\end{eqnarray}
 $\phi(x)$ is a positive, even function that is centered around $x = 0$; its  Fourier transform $\tilde{\phi}(p)$  centered around $p = 0$. The position and momentum spreads $\sigma_x$ and $\sigma_p$ of $\phi$ are assumed sufficiently small so that  $\sigma_p << k_0$ and
 $|x_0+a| >> \sigma_x$.  Then, $\tilde {\psi}_0(k) = \tilde{\phi}(k-p_0) e^{- i kx_0}$. By Eq. (\ref{fe+}),

\begin{eqnarray}
\psi_{\tau}(x) =   \int_0^{\infty}  dk \tilde{\phi}(k-p_0) e^{r_k(x) + i \theta_k(x) - i kx_0 - i \frac{k^2}{2m}\tau}  \label{propag}
\end{eqnarray}
We evaluate  the integral  in Eq. (\ref{propag}) using the saddle-point approximation: we set $r_k(x) \simeq r_{p_0}(x)$, $k^2 \simeq k_0^2 +  2k_0 (k - k_0)$,  and $\theta_k(x) = \theta_{k_0}(x) + \omega_{k_0}(x)(k-k_0)$, where
\begin{eqnarray}
\omega_{k}(x) = \frac{ \partial \theta_k(x)}{  \partial k}. \label{omegadef}
\end{eqnarray}
 Since $k_0 >> \sigma_p$, we can    extend the range of integration to $(-\infty, \infty)$ with insignificant error. Then,
\begin{eqnarray}
\psi_{\tau}(x) =  \phi\left[\omega_{k_0}(x) - x_0 - k_0\tau/m \right] e^{ r_{k_0}(x) + i \theta_{k_0}(x) -ik_0x_0 - i \frac{k_0^2}{2m}s}. \label{propag3}
\end{eqnarray}
 Thus,  $|\psi_{\tau}(x)|^2 =  e^{2 r_{k_0}(x) }\phi^2\left[\omega_{k_0}(x) - x_0 - k_0\tau/m \right]$. Since
  $\phi$ is peaked around zero, the maximum of $|\psi_{\tau}(x)|^2$ occurs for
\begin{eqnarray}
\omega_{k_0}(x) = x_0 + \frac{k_0}{m} \tau. \label{classicaleq}
\end{eqnarray}

 Eq. (\ref{classicaleq})  describes the correlation between position $x$ and time $\tau$ for particles that have exited the barrier. It has a restricted range of applicability when compared to Eq. (\ref{quasicl}). It applies only for potentials in which the saddle-point evaluation of Eq. (\ref{propag}) is a good approximation. This is not the case, for example,  for the double-well barrier \cite{KuMa}. Nonetheless, Eq. (\ref{classicaleq}) is expected to be a good approximation for $\cap$-shaped potentials that vary at a macroscopic scale.

If the first detector is placed in a classically allowed region, Eq. (\ref{classicaleq}) coincides with the  classical equations of motion. To show this, we employ the WKB eigenstates
$f_{k+}$ of the Hamiltonian
\begin{eqnarray}
f_{k+}(x) = \frac{C}{\sqrt[4]{k^2-2mV(x)]}} e^{i\int_{x_0}^x \sqrt{k^2-2mV(x')]}dx' + i kx_0},
\end{eqnarray}
 where   $C$ is a real constant and $x_0 < - a$.  We readily find that
 By Eq. (\ref{omegadef}),
 \begin{eqnarray}
 \omega_k(x) = \int_{x_0}^x dx' \frac{k}{\sqrt{k^2 - 2mV(x')}}           + x_0.
 \end{eqnarray}
 Hence, Eq. (\ref{classicaleq}) implies that  $\int_{x_0}^x  \frac{dx'}{\sqrt{k_0^2 - 2mV(x')}}   = \tau$,
 or equivalently,
 \begin{eqnarray}
 \frac{1}{2} m \left(\frac{dx}{d\tau}\right)^2 +V(x) = \epsilon_{k_0}. \label{encon}
 \end{eqnarray}

 However, the WKB approximation describes  the quasi-classical equations (\ref{classicaleq}) in the classically forbidden region very poorly. We show this in Sec. 4.2, where we compare the quasi-classical equations derived from the exact eigenstates of the Hamiltonian with the ones derived from the WKB approximation. The latter are constructed in the Appendix B.

\subsection{Tunnelling in the square barrier}
Next, we derive the quasi-classical equations (\ref{classicaleq}) for the square barrier potential
\begin{eqnarray}
V(x) = \left\{ \begin{array}{cc}  V_0 & -a \leq x \leq a \\0 &|x| > a.\end{array} \right.
\end{eqnarray}
 The eigenfunctions of the Hamiltonian $f_{k+}$ in the  forbidden region are
\begin{eqnarray}
f_{k+}(x) =  \frac{e^{ika}  T_k}{\sqrt{2\pi}} \left[ \cosh \lambda (a-x) - \frac{ik}{\lambda}     \sinh \lambda (a-x) \right], \label{eigenfb}
\end{eqnarray}
where $\lambda = \sqrt{2mV_0-k^2}$ and
$T_k$ is the transmission amplitude
\begin{eqnarray}
T_k = \frac{e^{-2ika}}{\cosh 2 \lambda a + \frac{i}{2} (\frac{\lambda}{k} - \frac{k}{\lambda})]\sinh 2 \lambda a  }.  \label{transmi1}
\end{eqnarray}

Writing $T_k = |T_k|e^{i \phi_k}$,    the associated phase function becomes
\begin{eqnarray}
\theta_k(x) = ka + \phi_k + \mbox{Im} \log \left[1 - i \frac{k}{\lambda} \tanh \lambda (a-x) \right].
\end{eqnarray}
Then,
\begin{eqnarray}
\omega_k(x) = a +\phi'_k + \frac{k}{m} \beta_k(x),
\end{eqnarray}
where
\begin{eqnarray}
\beta_k(x) = \frac{m}{k} \frac{a-x - \frac{2mV_0}{k^2\lambda} \sinh \lambda (a- x) \cosh \lambda(a-x)  }{\frac{2mV_0}{k^2} \cosh^2\lambda(a-x) -1}.
\end{eqnarray}

We define the shifted time parameter $\tau_1 = \tau + m\frac{x_0 +a}{p_0}$,  so that $\tau_1 = 0 $ the moment that the particle enters the barrier region. Then, Eq. (\ref{classicaleq}) becomes
\begin{eqnarray}
\tau_1 = t_{ph}(k_0) +  \beta_{k_0}(x),  \label{class2}
\end{eqnarray}
where
\begin{eqnarray}
t_{ph}(k) := \frac{m}{k} [ \phi'_k + 2a] = \frac{m}{k} \frac{\frac{1}{4\lambda} (\frac{\lambda}{k}+\frac{k}{\lambda})^2 \sinh 4\lambda a + (1 - \frac{k^2}{\lambda^2})a }{1 + \frac{1}{4} (\frac{\lambda}{k}+\frac{k}{\lambda})^2 \sinh^2 2\lambda a }
 \end{eqnarray}
 is the Wigner-Bohm {\em phase time} \cite{BW}, one of the candidates for the  tunneling time.

We use the dimensionless variables $D= x/a$ and $S= \frac{k_0}{m} \lambda \tau_1$, and  Eq. (\ref{class2}) becomes
\begin{eqnarray}
S = \frac{\gamma \epsilon (1-D) - \frac{1}{2} \sinh [2 \gamma(1-D)]}{\cosh^2\gamma(1-D) - \epsilon} + \frac{\frac{1}{4}(1+\epsilon) \sinh 4\gamma + (1-2\epsilon)\gamma }{1-\epsilon +\frac{1}{4} (1+\epsilon) \sinh^2 2\gamma}, \label{SD}
\end{eqnarray}
where  $\gamma = \lambda a$ quantifies the opacity of the barrier and
$\epsilon = \frac{k_0^2}{2mV_0} < 1$.

Interestingly,  $dS/dD \geq 0$ for all $D \in[-1,1]$ and for all   values of $\gamma$ and $\epsilon$. There is an one-to-one correspondence between $S$ and $D$, hence, Eq. (\ref{SD}) can be inverted, and  classical path $x(\tau)$   can be identified. The properties of those paths are the following.
\begin{enumerate}[(i)]
\item Since $\beta_k(a) = 0$,  a detector placed at the barrier's exit,   records $\tau_1 = t_{ph}(k_0)$ for all values of $\gamma$ and $\epsilon$. This is compatible with the idea that the phase time is the time it takes for a particle to cross the barrier, at least as far as time-of-arrival measurements are concerned \cite{AnSav13}.

\item  For a thin barrier ($\gamma << 1$), $S$ increases linearly with $D$:
\begin{eqnarray}
S = \gamma \left[ D+ (1-\epsilon)^{-1}\right].
  \end{eqnarray}
  This equation corresponds to  a classical free particle  traversing the forbidden region with constant velocity $dx/d\tau = k_0/m$, as though   the barrier is not present.

\item    For an opaque barrier ($\gamma >> 1$),   $S \simeq 1$ everywhere except for a neighborhood of width $\gamma^{-1}$ around $D = 1$, where $S$ increases  rapidly to   $2$ at $D = 1$. Far from the exit point $D = 1$,
 the velocity
    \begin{eqnarray}
    \frac{dD}{dS} \simeq \frac{1}{8 \gamma^2\epsilon (1-D)} e^{2\gamma (1-D)} \label{hart}
    \end{eqnarray}
 becomes arbitrarily large as $\gamma \rightarrow \infty$. This is another manifestation of the  Hartmann effect \cite{Har62}, i.e., the apparent possibility of a superluminal signal for extremely opaque barriers. This apparent superluminality is an artifact of the chain of approximations involved in deriving (\ref{hart}), and, in particular, the consideration of non-relativistic particles.
 The probability densities in the QTP method are obtained from the correlation functions of a local relativistic quantum field theory that fully respect causal propagation of signals \cite{AnSav13}.

\end{enumerate}
Indicative plots of $S$ as a function of $D$ are given in Fig. 1.

\begin{figure}[tbp]
\includegraphics[height=6cm]{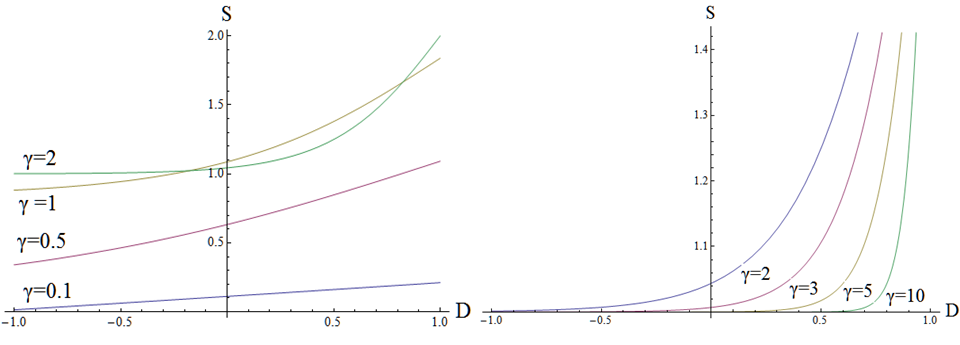} \caption{ \small  Plot of $S$ of  Eq. (\ref{SD}) as a function of $D$ for $\epsilon = 0.1$ and different values of the parameter $\gamma$.}
\end{figure}

We also compare the quasiclassical equations obtained from the exact eigenfunctions (\ref{eigenfb}) to the ones obtained from the WKB eigenfunctions. The latter equations are derived in the Appendix B. Expressed in terms of the $S$ and $D$ variables, the WKB quasiclassical equation (\ref{pathcorr}) for the square barrier is
\begin{eqnarray}
S = \frac{\gamma\sqrt{\epsilon^{-1}-1}(1-D)}{\cosh\left[ \log2 + 2 \gamma (1-D)\right]}  \label{SDwkb}
\end{eqnarray}

Eqs. (\ref{SD}) and (\ref{SDwkb}) define functions $S(D)$ with very different qualitative behavior; this can be seen in the plots of Fig.2. In particular, the function $S(D)$ of Eq. (\ref{SDwkb}) is not invertible and thus,  does not define classical paths. Furthermore, the WKB approximation completely misrepresents the behavior of $\tau_1$ near the exit point $x=a$, since it predicts $\tau_1 = 0$, while the exact solution gives $\tau_1 = t_{ph}(k_0)$. We conclude that the WKB approximation is unreliable for estimating the quasi-classical equations (\ref{classicaleq}) in the forbidden region. This is not unexpected, as the WKB approximation is also unreliable for estimating the tunneling times.

\begin{figure}[tbp]
\includegraphics[height=11cm]{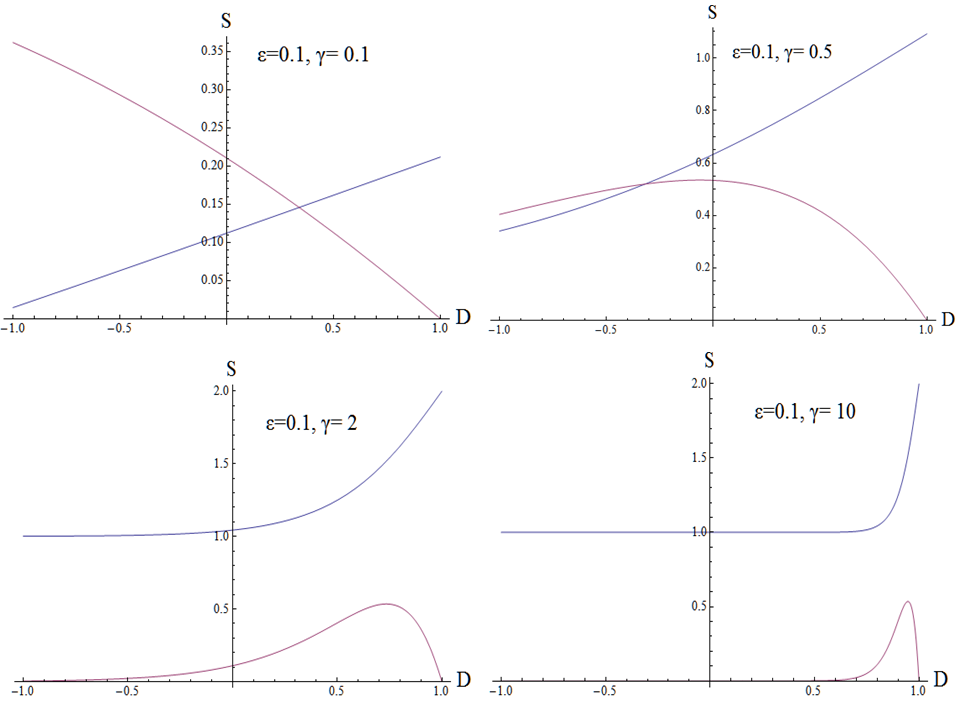} \caption{ \small  Comparison of the function $S(D)$  of Eq. (\ref{SD}) that is obtained from the exact eigenstates of the Hamiltonian to the function $S(D)$  of Eq. (\ref{SDwkb}) that is obtained from the WKB eigenstates. Each plot corresponds to a different choice of the dimensionless parameters $\gamma$ and $\epsilon$. }
\end{figure}

 \section{Conclusions}
We have identified quasi-classical equations relating time $\tau$ and position $x$ for a particles tunneling through a classically forbidden region. It was obtained by maximizing the post-selected probability density $P_{p.s.}(x, \tau)$ that a particle that eventually exits the barrier has been recorded at time $\tau$ by a detector located at $x$ in the classically forbidden region. The probability density $P_{p.s.}(x, \tau)$ was constructed by considering sequential time-of-arrival measurements using the QTP method.

Tunneling paths derived by our method do not involve any complexification and they have a precise operational significance. They are defined in terms of probabilistic correlations that are, in principle, measurable.   Moreover, the quasi-classical equations do not depend on any properties of the apparatus, they are expressed solely in terms of the solution to Schr\"odinger's equation. This suggests that they may not be tied to the specific measurement set-up considered here, but may arise  in more complex schemes.

 Our results demonstrate the possibility of writing  classical equations for quantum systems, even in absence of a correspondence to classical physics. The key point is the quantum system is characterized by strong  probabilistic correlations \cite{GeHa}.  In fact, the paths derived here are not accessible by the usual semi-classical methods. Thus, our method is of direct interest to fields where new notions of emergent classicality may be required, in particular, in the study of macroscopic quantum phenomena and in quantum cosmology \cite{GeHa94}.

\newpage

\begin{appendix}

\section{Detection and non-detection events in the joint time-of-arrival measurement}
Eq. (\ref{pdouble}) describes the joint time-of-arrival probability distribution in two detectors. In this section, we describe how it defines a properly normalized probability distribution by including the events that no particle has been recorded in either detector. Thus, we estimate how the first time-of-arrival measurement affects the total transmission rate through the barrier.

We evaluate the constant $C$ by choosing a convenient normalization condition. To this end, we consider the case of free particles, studied in \cite{AnSav16}.  We substitute $T_k=1, R_k=0$, and $f_k(x) = e^{ikx}/\sqrt{2\pi}$.

For an initial state with strictly positive momentum content, the probability density (\ref{pdouble}) is strongly suppressed   for $\tau < 0$ or if $t < \tau$. Hence, the total probability $\mbox{Prob}(x, L)$ that two detection events have occurred is well approximated by integrating both $\tau$ and $t$ along the full real axis,

 \begin{eqnarray}
 \mbox{Prob}(x, L) := \int_{-\infty}^{\infty} d\tau \int_{-\infty}^{\infty}dt  P(x,s;  L, t)
 \nonumber \\
 = Cm^2  \int \frac{dk_1 dk_2}{2\pi k_1 k_2} \tilde{g}_1(\epsilon_{k_1} - \epsilon_{k_2}) \tilde{g}_2(\epsilon_{k_2}) \theta(k_2) |\tilde{\psi}_0(k_1)|^2. \label{prl1l2}
 \end{eqnarray}
Note that if the probabilities involve smearing also with respect to position, then $\tilde{g}_1$ in (\ref{prl1l2}) has also a dependence on $k_1 - k_2$. Thus,  $\tilde{g}_1$ is a function of two variables, which can be chosen as $\epsilon_{k_1}$ ans $\epsilon_{k_2}$. Hence, in Eq. (\ref{prl1l2}), we   write $\tilde{g}_1(\epsilon_{k_1}, \epsilon_{k_2})$ rather than $\tilde{g}_1(\epsilon_{k_1}-\epsilon_{k_2})$. This point is important for an exact definition of the limit of   ideal detectors.

 Eq. (\ref{prl1l2}) implies that $\tilde{g}_2(\epsilon_k)/|k|$ is the absorption coefficient of the second detector, while $\frac{1}{|k_1|}\tilde{g}_1(\epsilon_{k_1}, \epsilon_{k_2})$ is the probability that an incoming particle of momentum $k_1$ is scattered to a different momentum $k_2$. We define ideal detectors as follows.

 For the first detector, we assume that energy
  transfer during scattering   is negligible. This implies that $\tilde{g}_1(\epsilon_{k}, \epsilon_{k'}) \sim \epsilon_k \delta (\epsilon_k - \epsilon_{k'})$. For the second detector, we assume, as in Sec. 2.2, that particle absorption is independent of the particle's momentum. Hence,
    t $g_{2}(\epsilon) \sim \sqrt{\epsilon}$. We choose the constant $C$ so that  $\mbox{Prob}(L_1, L_2) = 1$, hence, we obtain Eq. (\ref{Fker3}).

  Next, we proceed to the definition of probabilities associated to the event of no detection. The general procedure is described in Ref. \cite{AnSav15}.  We note that any probability assignment for the time of arrival must also include probabilities for the event of no arrival, otherwise the set of alternatives is not exhaustive \cite{AnSav06}. In what follows, we will denote the event of detection  by $+$. The event of no detection will be denoted by $\emptyset$.

  Additivity of probabilities implies
  \begin{eqnarray}
  P(x,\tau;L,+) + P(x ,\tau, L, \emptyset) = P_1(x,\tau),
  \end{eqnarray}
  where
  \begin{itemize}
  \item $P(x, \tau;L,+)$ is the probability density that a particle was recorded at the first detector at time $\tau$ and then recorded at the second detector. $P(x, \tau, L,+)$ is  the post-selected probability density (\ref{pps}).
  \item     $P(x, \tau;L,\emptyset )$  is the probability density  that a particle was recorded at the first detector at time $\tau$ but not recorded at the second detector.
      \item $P_1(x, \tau)$ is the probability density  that the particle is detected in the first detector irrespective of what happens later. Since no probability density should be affected by the outcomes of any subsequent measurement,  it is given by Eq. (\ref{plt1}). Using the same normalization condition as above, we obtain
          \begin{eqnarray}
          P_1(x, \tau) = \int dk dk' \sqrt{\frac{\epsilon_k+\epsilon_{k'}}{m}} f_k(x)f_{k'}(x) e^{-i(\epsilon_k-\epsilon_{k'})\tau} \tilde{\psi}_0(k) \tilde{\psi}^*_0(k')
          \end{eqnarray}
  \end{itemize}

  Additivity also implies that

   \begin{eqnarray}
  P(x, +; L, t) + P(x ,\emptyset, L, t) = P_2(L, t),
  \end{eqnarray}
  where
  \begin{itemize}
  \item $P(x, +; L, t)$ is the probability density  that a particle was recorded at some time at the first detector and then recorded at the second detector at time $t$. It is given by the integral $\int_0^{t} d\tau P(x,\tau; L,t)$. Since for $\tau< 0$ and $\tau > t$, $P(x, +; L, t)$ is suppressed, we can extend the integration over $\tau$ to the full real axis. By Eqs. (\ref{pdouble}) and (\ref{Fker3}), we obtain
      \begin{eqnarray}
        P(x, +; L, t)  = 4 \sqrt{m} \int \frac{dk dk'}{kk'} (\epsilon_k +\epsilon_{k'})^{3/2} e^{i(k-k')L - i (e\epsilon_{k}-\epsilon_{k'})t} f_k(x)f^*_{k'}(x)
        \nonumber \\
  \times \left[ T_k f^*_{k+}(x) + R_kf^*_{k-}(x)\right] \left[ T^*_{k'} f_{k'+}(x) + R^*_{k'} f_{k-}(x)\right]
      \end{eqnarray}
      \item $P(x, \emptyset; L, t)$ is the probability density  that a particle was not at the first detector and but was recorded at the second detector at time $t$. Since there is no record at the first detector, the probability is simply given by (\ref{probden2}).
          \item $P_2(L, t)$ is the probability density  that the particle was recorded at the second detector irrespective of what happened at the first detector.
  \end{itemize}

 From the expressions above, we  define the following integrated probabilities
 \begin{enumerate}
 \item $P_{++}: = \int_0^{\infty} dt P(x, +; L, t) = \int_0^{\infty} d \tau P(x, \tau;L,+)$ is the probability that the particle has been recorded by both detectors. We find
     \begin{eqnarray}
P_{++} = 2\pi \int_0^{\infty} dk  |f_{k+}(x)|^2 |T^*_k f_{k+}(x) + R^*_kf_{k-}(x)|^2 |\tilde{\psi}_0(k)|^2.
\end{eqnarray}
\item $P_{+\emptyset} := \int d \tau(x, \tau; L,- )$ is the probability that the particle was recorded at the first but not at the second detector. We find
\begin{eqnarray}
P_{+\emptyset} = \int dk  |f_k(x)|^2 |\tilde{\psi}_0(k)|^2 - P_{++}
\end{eqnarray}
where ${\cal F}^+(x, E)$ is given by Eq. (\ref{Fkint2}).
\item $P_{\emptyset +} := P(x, \emptyset; L, t)$ is the probability that the particle was recorded at the second but not at the first detector. We find
    \begin{eqnarray}
    P_{\emptyset +}  = \int dk |T_k|^2 |\tilde{\psi}_0(k)|^2.
    \end{eqnarray}
\item $P_{\emptyset\emptyset} := 1 -     P_{++} - P_{+\emptyset}  - P_{\emptyset +}$ is the probability that neither detector recorded the particle.

 \end{enumerate}

 For an almost monochromatic initial state with mean momentum $k_0$, the probabilities above become
 \begin{eqnarray}
 P_{++} &=&   |f_{k_0+}(x)|^2 |T^*_{k_0} f_{k_0+}(x) + R^*_{k_0}f_{k_0-}(x)|^2\\
 P_{+\emptyset} &=&  |f_{k_0+}(x)|^2 \left[ 1 - |T^*_{k_0} f_{k_0+}(x) + R^*_{k_0}f_{k_0-}(x)|^2\right]\\
 P_{\emptyset +} &=& |T_{k_0}|^2.\\
 P_{\emptyset\emptyset} &=& 1 -     P_{++} - P_{+\emptyset}  - P_{\emptyset +}.
 \end{eqnarray}
In opaque barriers $P_{++} << P_{+\emptyset} <<  P_{\emptyset\emptyset}$ and $P_{++} <<  P_{\emptyset +} <<  P_{\emptyset\emptyset}$. Particles recorded by both detectors are only a small fraction  of particles recorded by both.

\section{Quasi-classical equations in the WKB approximation}
Here we derive the quasi-classical equations (\ref{classicaleq}) when tunneling is described in the WKB approximation. As shown in Sec. 4.2 the WKB approximation is completely unreliable in the tunneling region.

We consider a particle tunneling through a barrier potential $V(x)$, as described in Sec. 4.1. The forbidden region corresponds to the interval  $(-x_1, x_1) \subset (-a, a)$, where $x_1$ is  the positive solution of the equation $E= V(x_1) $.  The WKB eigenfunctions $f_{k+}(x)$ for $x \in (-x_1, x_1)$ are
 \begin{eqnarray}
 f_{k+}(x) =\frac{1}{\sqrt{2\pi \lambda(x)}} \frac{e^{i \int_{-a}^{-x_1} dx \sqrt{k^2-2mV(x)]}-ika +i\frac{\pi}{4} }}{  \frac{1}{2}e^{-\int_{-x_1}^{x_1} dx' \lambda(x')}+ 2e^{\int_{-x_1}^{x_1} dx' \lambda(x')}} [e^{-\int_x^{x_1} dx' \lambda(x')} -2 i e^{\int_x^{x_1} dx' \lambda(x')}],
 \end{eqnarray}
 where $\lambda(x) = \sqrt{2mV(x) - k^2}$.

 The phase $\theta_k(x)$ is
 \begin{eqnarray}
 \theta_k(x) = \int_{-a}^{-x_1} dx \sqrt{k^2 - 2mV(x)}-  ka + \frac{\pi}{4} + \mbox{Im} \log \left( 1 - 2i  e^{2\int_{x}^{x_1} dx' \lambda(x')}\right).
 \end{eqnarray}
By Eq. (\ref{omegadef}),
 \begin{eqnarray}
 \omega_k(x) = \int_{-a}^{-x_1} \frac{k/m}{\sqrt{k^2-2mV(x)}}dx - a + \frac{k}{m} \beta_k(x),
  \end{eqnarray}
  where
  \begin{eqnarray}
\beta_k(x) =   \frac{m  \int_x^{x_1} \frac{dx'}{\sqrt{2mV(x') - k^2}}}{ \cosh \left[ \log 2+ 2\int_{x}^{x_1} dx' \sqrt{2mV(x') -k^2}\right]}.   \label{betak}
 \end{eqnarray}

 It is convenient to express the correlation equation (\ref{classicaleq}) in terms of the shifted time variable
 \begin{eqnarray}
 \tau_1 = \tau - t(x_0, p_0)
 \end{eqnarray}
 where
 \begin{eqnarray}
 t(x_0, p_0) =  \int_{-a}^{-x_1}  \frac{dx'}{\sqrt{k_0^2 - 2mV(x')}} - \frac{m(x_0+a)}{p_0},
 \end{eqnarray}
 is the time it takes a classical particle starting at $x = x_0 < - a$ and with momentum $k_0$ to arrive at the turning point $x = - x_1$. Hence, $\tau_1$
 is zero when the  particle enters the forbidden region.

 Then, Eq. (\ref{classicaleq})   becomes
 \begin{eqnarray}
 \tau_1 =  \beta_{k_0}(x) \label{pathcorr}
 \end{eqnarray}

By Eq. (\ref{betak}), $\tau_1$ is always positive and  has an upper bound
$\tau_1 \leq \frac{4m}{5} \int_{-x_1}^{x_1} \frac{dx'}{\sqrt{2mV(x') - k^2}}$.
Furthermore, $\tau_1$ vanishes for $x = x_1$ but is non-zero at $x = -x_1$. If Eq. (\ref{pathcorr}) were interpreted in terms of classical paths, this would mean that the detection at the exit of the barrier
 takes place earlier than a detection just before the entry of the barrier region, and that there    exists a discontinuity in time as the particle enters the classically forbidden region.

\end{appendix}

 \end{document}